\documentclass[showpacs,twocolumn,amsmath,amssymb,fixfloats]{revtex4}
\usepackage[english]{babel}
\usepackage{amsmath}
\usepackage{amssymb}
\usepackage{graphicx}

\newcommand{\ket}[1]{|#1\rangle}

\begin{document}

\title{Spin squeezing in optical lattice clocks via lattice-based QND
    measurements}

\author{D. Meiser}
%\email{dmeiser@jila.colorado.edu}
\author{Jun Ye}
\author{M. J. Holland}
\affiliation{JILA, National Institute of Standards and Technology
and University of Colorado, Boulder, CO 80309-0440, USA}

\begin{abstract}
Quantum projection noise will soon limit the best achievable precision of
optical atomic clocks based on lattice-confined neutral atoms.
Squeezing the collective atomic pseudo-spin via measurement of the
clock state populations during Ramsey interrogation suppresses the
projection noise. We show here that the lattice laser field can be used to
perform ideal quantum non-demolition measurements without clock shifts or
decoherence and explore the feasibility of such an approach in
theory with the lattice field confined in a ring-resonator.
Detection of the motional sideband due to the atomic vibration in
the lattice wells can yield signal sizes a hundredfold above the
projection noise limit.
\end{abstract}

%\pacs{42.62.Eh; 32.80.-t; 32.80.Qk; 42.62.Fi}
\maketitle

\section{Introduction}

Optical atomic clocks based on neutral atoms confined in
state independent optical lattices have made dramatic progress
recently \cite{Takamoto:LatticeClock,ludlow:033003,LeTargat1}. The highest spectral
resolution has been achieved in such a system \cite{Boyd1}, resulting in the
clock instability approaching $1\times10^{-15}$ at 1 s and an overall
uncertainty reaching $1\times10^{-16}$ \cite{Boyd2,Ludlow2}. The signal to
noise ratio (SNR) for $10^4$ atoms in these latest experiments is within a
factor of two of the quantum projection noise limit \cite{Ludlow2}. As the
lattice clock performance approaches this limit squeezing of the collective
atomic pseudo-spin to overcome quantum projection noise
\cite{PhysRevA.46.R6797,PhysRevLett.82.4619} will lead to dramatic further advances in the clock
performance because the number of atoms involved is large. Furthermore, we
believe that the precision, control, and isolation from the environment
achieved in these metrological lattice systems can be the basis of powerful
probes to explore novel quantum dynamics, such as manifested in the collective
interactions between an atomic ensemble and an optical cavity
\cite{Vuletic:CollectiveCooling,Vuletic:CollectiveFriction,
    Kruse:AtomsInRingCavity,Nagorny:Collective2003,Hemmerich,
    Esslinger:BECinCavity,Horak:Dissipative_BEC_cavity,Domokos:Mechanical_effects_light,
    mekhov:100402}. 

\begin{figure}
\includegraphics[width=8cm]{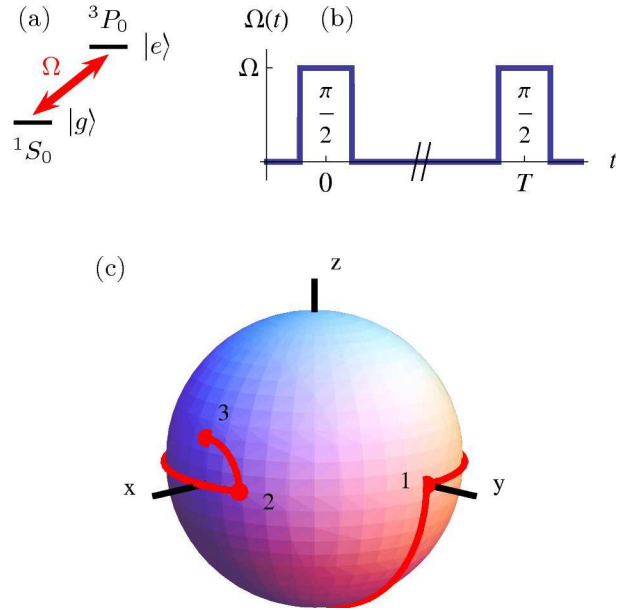}
\caption{(Color online) (a) $\ket{g}$ and $\ket{e}$ are the two
clock states for $^{87}$Sr. (b) The Ramsey pulse sequence with
each pulse having an area of $\pi/2$, and they are separated by a free
evolution time $T$ much longer than the duration of each pulse. (c)
Illustration of the trajectory of the atomic pseudo spin on the Bloch sphere
during the Ramsey pulse sequence. 1: action of the first pulse, 2: free evolution, and 3: action of the second pulse. After the second pulse the population difference of the clock states is measured.}
\label{clockschematic}
\end{figure}

An atomic clock can be realized with the Ramsey technique illustrated in
Fig. \ref{clockschematic}(a-c): atoms with two clock states $\ket{g}$ and
$\ket{e}$ are driven with two $\pi/2$-pulses separated by a free evolution time
$T$. The evolution of the atomic state during this clock sequence (Fig.
\ref{clockschematic}(b)) can be visualized on the Bloch sphere (Fig.
\ref{clockschematic}(c)): The atomic pseudo spin points initially toward the
south pole and is rotated around $x$ by the first $\pi/2$-pulse to lie along
$y$ at position 1. From there the pseudo spin precesses along the equator to
reach position 2 after $T$. This precession is the ``ticking'' of the atomic
clock: The total angle by which the pseudo-spin precesses measures the elapsed
time. In order to measure this phase angle the position of the Bloch vector in
the equatorial plane has to be translated into a position in the $x$-$z$-plane
by the second $\pi/2$-pulse to position 3 where it can be measured through the
population difference between the levels which corresponds to the $z$-component
of the pseudo-spin. 

\begin{figure}
\includegraphics[width=8cm]{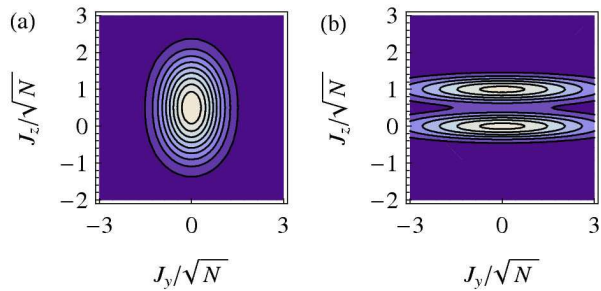}
\caption{(Color online) Sum of probability distributions at the end of the
    clock sequence (position 3 in Fig.~\ref{clockschematic}(c)) for two pseudo
	spin vectors that we wish to resolve in order to distinguish two
	atom-field detuning values. The pseudo spins have accumulated a
	relative phase $1/\sqrt{N}$ during the free evolution stage of the
	Ramsey sequence. In (a) the initial state of the atoms was the
	conventional state of all atoms in $\ket{g}$ and the two final
	positions of the pseudo spin cannot be distinguished, giving rise to
	the shot-noise limit. In (b) on the other hand a spin squeezed
	state was used and the two positions can be resolved leading to a
	measurement precision better than the
	shot-noise limit.}
\label{squeezed}
\end{figure}

The projection noise originates in the tip of the collective pseudo-spin not
pointing in a sharp direction. Rather the position of the tip of the
pseudo-spin is distributed with a width of order $\sqrt{N}$ for $N$ independent
atoms. The ``hand'' of the atomic clock is intrinsically fuzzy. As a
consequence two phase angles closer to each other than $1/\sqrt{N}$ cannot be
distinguished. This is illustrated in Fig.~\ref{squeezed} where we show the sum
of the probability distributions of two Bloch vectors that we wish to
distinguish. In Fig.~\ref{squeezed}(a) the two Bloch vectors have accumulated
a relative phase of $1/\sqrt{N}$ at the end of the Ramsey pulse
sequence at position 3 and hence the two Bloch vectors are not resolved. 

However, the projection noise can be reduced by preparing the atoms in an
entangled state in which the distribution of the pseudo spin is squeezed.  The
principle of this idea is to prepare the atoms in a state such that the
distribution of the pseudo spin at the end of the clock sequence is narrower in
the $z$-direction, as shown in Fig. \ref{squeezed}(b). This way the Bloch
vectors of two states that have accumulated less than $1/\sqrt{N}$ of phase
difference can be distinguished. The higher phase resolution translates into an
increased clock precision. By means of spin squeezing the projection noise can
be reduced in principle to the Heisenberg limit which scales with the number of
atoms as $1/N$. The Heisenberg limit scaling has been demonstrated in
experiments with few entangled ions \cite{Leibfried,Roos}. For neutral
atoms, their large sample number permits a huge reduction of the projection
noise \cite{Kuzmich,Polzik}.

Spin-squeezed states can be created by means of atom interactions 
\cite{C.Orzel03232001}, interaction with a cavity field \cite{Molmer,PhysRevA.62.063812} or through
the back-action of quantum non-demolition (QND) measurements
\cite{Polzik,JMGeremia04092004,chaudhury:043001}. Since we are considering the latter approach in
this paper we briefly review this type of measurement. For definiteness we
discuss the case of measurements of the $z$-component $\hat J_z$ of the
collective atomic pseudo spin. This is no restriction since  squeezed states
along any direction can be obtained from a state squeezed along the
$z$-direction through rotations.

According to the principles of quantum mechanics we are guaranteed to get the
same result in a second measurement of any observable as in a first measurement
of that observable provided that the observable commutes with the Hamiltonian.
This can be considered an extreme case of squeezing: The probability
distribution of the observable has collapsed to a delta-function due to the
first measurement. This type of measurement is sometimes called a projective
measurement. The measurements that are of interest to us in this article are
weaker than projective measurements in the sense that several atomic states
corresponding to different eigenvalues of the observable are consistent with
the measurement outcome. The key point is that, conditioned on the measurement
record, some states become more likely than others. This has qualitatively
similar consequences for subsequent measurements of the observable as in the
projective measurement case. As we keep measuring the observable we are more
and more certain about the measurement outcome because only a small range of
results will be consistent with the measurement record obtained up to that
time. It is essential that these measurements be performed on the same system
and not on different copies. Therefore the measurement has to be
non-destructive. This type of measurement in which information about the state
of the system is extracted gently in such a way that the state of the system
persists after the measurement is complete is called a non-demolition
measurement. As shown above the resulting state is squeezed.

In our case the observable $\hat J_z$ is a collective variable. Therefore $\hat
J_z$ cannot be measured through a  measurement of each atom's spin, if we wish
to achieve spin squeezing. That would project the atoms on a product state in
which each atom's spin is pointing either up or down. Such a state has a
reduced total angular momentum $\langle \hat J^2\rangle$ and the reduction of
the length of the angular momentum outweighs any possible benefits from
squeezing. A measurement scheme for spin squeezing must therefore ensure that
the total angular momentum $\hat J^2$ is conserved. Mathematically speaking
this means that the Hamiltonian and the interactions describing the measurement
must commute with $\hat J^2$. Physically it means that inhomogeneous broadening
has to be negligible during the time scales of interest and the measurement
must not be able to distinguish between different particles.

To summarize, a measurement protocol for the preparation of spin-squeezed
states should satisfy the following requirements. First, the probe must not
lead to decoherence by spontaneous emission or depolarization of the atomic
sample by inhomogeneous effects (non-demolition and conservation of $\hat J^2$
requirements). Second, the measurement must give the population difference with
precision exceeding the atomic projection noise. Finally, for clock
applications it is important that the measurement does not introduce shifts of
the clock transition.

In this article we consider neutral atoms in an optical lattice clock.
Conventionally the lattice is treated as an external potential in these
systems, i.e. the back-action of the atoms on the light field is neglected. This
approximation is motivated by the coupling between atoms and lattice photons
being very weak. A large number of photons is necessary to provide a deep
enough lattice for the atoms while the many orders of magnitude smaller number
of atoms has only a microscopic effect on the light field. However, the minute
changes that the lattice fields experience when they propagate through the
atomic sample contain information about the atomic state that is normally lost.
We show that the information that the atoms imprint on the lattice fields can be
harnessed. In particular we propose to use the lattice field itself for the
non-demolition measurement of the clock pseudo-spin to achieve spin squeezing.
Such a scheme has several advantages over probing the atomic state with an
additional interrogation field in addition to using a resource that is normally
wasted. Importantly, the lattice does not introduce clock shifts as it operates
at the magic frequency where the two clock states have an identical
polarizability \cite{Takamoto:LatticeClock,ludlow:033003,LeTargat1,Boyd1,Boyd2,Ludlow2}.
Decoherence by spontaneous emission is small since the lattice is far detuned
from strong atomic transitions. The lattice laser also couples equally to atoms
at different lattice sites due to the lattice periodicity, i.e. the probe is
only sensitive to the total pseudo spin and the measurement is of the
non-demolition type. According to our list of requirements above it remains to
be shown that sufficient precision for spin squeezing can be achieved using
this approach. That is the main subject of this article.

In general one has to ``level the playing field'' between photons and atoms in
order for the microscopic effects of the atoms on the light field to become
detectable experimentally. Since it is difficult to increase the number of
atoms this means that the number of photons has to be reduced. This can be
achieved by putting the photons in a cavity. Figuratively speaking each photon
passes through the atomic cloud many times and therefore a much smaller number
of photons is sufficient to generate the optical lattice. Conversely each photon
also accumulates the effect of interaction with the atoms over many round trips
so that the signal is enhanced. In this article we consider the case of a bad
cavity, in a sense that we will make precise below, for two reasons. First,
this is the case that is immediately relevant for the next generation of
atomic clocks. Second, we want to use the light field as a measurement
device.  This implies that the dynamics of the field should be as simple as
possible so that the state of the atoms can be read off directly 
without having to understand the complicated physics of the meter. In a high-Q
cavity where the atoms and light field interact with each other in the strong
coupling regime their coupled dynamics would be so complicated that it would
become hard if not impossible to infer the atomic state from measurements of
the field.

The principle of our idea for measuring the atoms' spin with the lattice fields
is the following. The atoms are initially prepared in state $\ket{g}$.
During the first Ramsey pulse the clock laser drives them into a 50/50
superposition. If this drive is in the Lamb-Dicke and resolved side-band regime
the atoms will remain at rest. The recoil momentum associated with each
transition has to be taken up by the lattice fields. If the lattice is
generated in a ring cavity this is achieved by transferring photons from one
mode to the counter propagating mode. By measuring the intensity redistribution
one can determine how much momentum has been exchanged between the two modes
or, equivalently, how many transitions have happened. This constitutes a
measurement of $\hat J_z$ since we started from a state with a known $\hat J_z$
quantum number. 

The rest of this article is organized as follows. In section \ref{model} we
introduce the model and develop the theoretical framework that we
will use. Section \ref{IntensityImbalance} discusses the measurement scheme
outlined in the previous paragraph. As we will see, the signal to noise ratio
(SNR) of this measurement is insufficient to achieve spin squeezing for
currently realizable lattice clocks. Section \ref{SidebandSpectroscopy} is
dedicated to a superior  measurement scheme based on detecting motional
sidebands of the atoms that promises to lead to a strong enough signal that can
yield significant spin squeezing. We draw conclusions in section
\ref{Conclusion}.

\section{\label{model} Model}

\begin{figure}
\includegraphics[width=8cm]{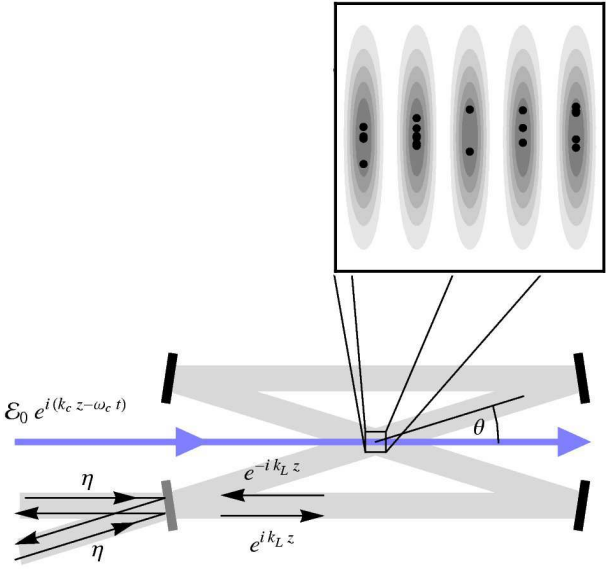}
\caption{(Color online) Schematic of the lattice in the ring resonator with
    atoms trapped at the intensity anti-nodes. The clock laser propagates along
	the $z$ axis and the projection of the wave numbers of the counter
	propagating lattice modes along the $z$-axis is $k_L$ and $-k_L$. The
	lattice light leaking out of the resonator contains information about
	the atomic state. The clock laser wave vector is commensurate with the
	reciprocal lattice period under this particular lattice geometry.
	Details of coupling into and out of the ring cavity as well as
	the detector arrangement are given in
	Fig.~\ref{detectionschemes}.}
\label{schematic}
\end{figure}

We consider $N$ two-level atoms with ground state $\ket{g}$, excited state
$\ket{e}$, and transition frequency $\omega_a$, trapped in a one-dimensional
optical lattice generated by the two counter propagating running wave modes of
a ring cavity with frequency $\omega_L$ (Fig. \ref{schematic}). The projections
of the running waves' wave vectors along the $z$-axis are $k_L=\cos \theta
\; \omega_L /c$ and $-k_L$, where $\theta$ is the angle at which clock laser and
the lattice beams cross. The transverse profile of the modes is approximately
Gaussian and we neglect the dependence of its radius $w_0$ on $z$. The cavity
length is $L$.  The atomic transition is probed by a highly stabilized clock
laser of frequency $\omega_c$, which is linearly polarized in the same
direction as the lattice.

\subsection{Effective Hamiltonian}

In our calculations, we neglect inhomogeneous effects. This is justified if the
duration of the clock pulse sequence is smaller than the $T_2$ time
associated with the inhomogeneities. The main source of inhomogeneous
broadening in experiments is due to the radial distribution of atoms in
the 1D optical lattice which is typically only weakly confining in the
transverse direction. The inhomogeneities stemming from this can be eliminated
by confining the atoms in a 3D lattice. We assume that the wavefunction of the
atoms can be factorized into a part for spin and $z$ on the one hand and all
other degrees of freedom on the other hand. Since we will be only interested in
spin and $z$, considering just that part of the wavefunction is sufficient,
leading to an effectively 1D theory. The initial state, all atoms in $\ket{g}$
and vibrational ground state \footnote{We will make more precise below in what
    sense all atoms are in the vibrational ground state.},
    \begin{equation}
    \ket{\psi(t=0)}=\underbrace{\ket{g,n=0;\;g,n=0;\ldots;\;g,n=0}}_{N\; \rm times},
    \label{initialstate}
    \end{equation}
is completely symmetric under particle exchange with respect to the pseudo-spin
and $z$ which are the degrees of freedom relevant to us. The system will
remain in the totally symmetric subspace since inhomogeneities are negligible.
Hence we can use a totally symmetrized basis and going over to the usual second
quantized formalism we can effectively treat the atoms as bosons.

The system can be described with the Hamiltonian
\begin{widetext}
\begin{eqnarray}
\hat H&=&\frac{-\hbar^2}{2M}\sum_{\sigma=g,e}\int dz\;  \hat \psi_\sigma^\dagger(z)\frac{\partial^2}{\partial_z^2} \hat \psi_\sigma(z)
   +
   \hbar g_0\sum_{\sigma=g,e}\int dz\; (e^{-ik_Lz}\hat a_{k_L}^\dagger+e^{ik_Lz}\hat a_{-k_L}^\dagger)(e^{ik_Lz}\hat a_{k_L}+e^{-ik_Lz}\hat a_{-k_L})\hat \psi_\sigma^\dagger(z)\hat \psi_\sigma(z)\nonumber\\
   &&+
   \frac{\hbar \Omega}{2} \left(e^{-i\delta t}\int dz\;e^{ik_cz}\hat\psi_e^\dagger(z)\hat\psi_g(z)+H.C.\right) +\sum_{p=\pm k_L} \left(-\hbar \Delta \hat a_p^\dagger\hat a_p+\hbar \eta (\hat a_p^\dagger+\hat a_p)\right).\label{Hamiltonian}
\end{eqnarray}
\end{widetext}
$\hat \psi_e$ and $\hat \psi_g$ are bosonic field operators
describing the atoms in the excited and ground states.
The first term in the Hamiltonian is the
atomic kinetic energy, with $M$ being the atomic mass. The second
term describes the lattice potential. $g_0=\frac{\alpha
\omega_L}{\pi \epsilon_0 w_0^2 L}$ is the coupling constant between
atoms and the lattice field. $\alpha$ is the common polarizability
of $\ket{g}$ and $\ket{e}$. $\hat a_{\pm k_L}$ are bosonic field
operators for the running wave modes of the ring resonator. The
third term describes the drive of the atomic transition by the clock
laser with a Rabi frequency $\Omega$ and detuning
$\delta=\omega_c-\omega_a$. The last term represents the detuning
$\Delta$ of the lattice laser from the resonator mode and the pump
of the lattice field with amplitude $\eta$, which is assumed to be
real. The losses of both modes with decay constant $\kappa$ can be described by means of a standard
Born-Markov Liouville operator
\begin{equation}
\hat L [\hat \rho]=-\frac{\hbar \kappa}{2} \sum_{p=\pm k_L}(\hat a_{p}^\dagger \hat a_{p}\hat \rho +\hat \rho\hat a_{p}^\dagger \hat a_{p}-2\hat a_{p}\hat \rho\hat a_{p}^\dagger),
\end{equation}
where $\hat \rho$ is the system density operator. We neglect spontaneous
emission from $\ket{e}$ which is justified because the lifetime of that level
is many times longer than the clock sequence considered here.

The periodic arrangement of atoms in the lattice strongly couples the two
counter propagating modes with a coupling frequency of order $g_0 N$. The
symmetric cosine and antisymmetric sine modes described by the field operators
\begin{equation}
\hat b_\pm=\frac{1}{\sqrt{2}}\left(\hat a_{k_L}\pm\hat a_{-k_L}\right)
\end{equation}
on the other hand are uncoupled for a perfect atomic lattice and weakly coupled
if the density distribution is slightly perturbed. Since it is exactly these
small deviations of the atomic density distribution from the perfect lattice
caused by the clock laser
that we wish to detect, we transform the lattice field to the symmetric
and antisymmetric modes $\hat b_+$ and $\hat b_-$. 
Because the cavity pumping amplitude $\eta$ is real only the
symmetric mode is pumped. In steady state the symmetric mode $\hat b_+$ has
therefore a large mean amplitude $\beta_+$ which gives rise to the lattice.
The steady state amplitude can be found from the equations of motion
\begin{eqnarray}
i\frac{d\langle \hat b_+\rangle}{dt}&=&-(\Delta +i\frac{\kappa}{2})\langle \hat b_+\rangle +g_0\int dz\sum_\sigma\langle\hat \psi_\sigma^\dagger (z)\hat \psi_\sigma(z)\rangle\times\nonumber\\
	&&\left(2\cos^2k_L z\; \langle \hat b_+\rangle+i\sin 2k_L z \;\langle \hat b_-\rangle\right) + \sqrt{2}\eta,
\label{aplusequation}
\end{eqnarray}
where we have used that the expectation values like $\langle\hat b_+\hat
\psi_\sigma ^\dagger\hat\psi_\sigma\rangle$ etc. factorize in steady state to a
good approximation. We find $\beta_+$ by setting $d\langle \hat
b_+\rangle/dt=0$,
\begin{equation}
\beta_+=\frac{\sqrt{2}\eta}{\Delta+i\frac{\kappa}{2}-2 g_0 \mathcal{C}(0)},
\end{equation}
where we have used 
$\langle \hat b_-\rangle=0$ in steady state and we have introduced
\begin{equation}
\mathcal{C}(t)=\int dz\sum_\sigma \cos^2k_L z \langle
\psi_\sigma^\dagger(z)\psi_\sigma(z)\rangle.
\end{equation}
We eliminate $\beta_+$ with the canonical transformation $ \hat b_+\rightarrow
\hat d_++\beta_+,\; \hat b_-\rightarrow \hat d_-$. $\beta_+$ gives rise to
an optical lattice of depth $2\hbar |\beta_+|^2 g_0$. Assuming a deep lattice,
we can neglect tunneling between different lattice sites and approximate the
lattice with a harmonic potential at each site. 

The atom-lattice coupling is trivially lattice periodic. The clock-laser-atom
interaction can also be made lattice periodic by crossing the lattice modes at an angle
such that $k_c=2k_L$. Atoms on different sites are
then indistinguishable and by transforming the coordinates of a particle
trapped on site $j$ according to $z\rightarrow z-j\pi/k_L$ we can treat them as
if they were all trapped in a single harmonic well.  This interpretation makes
precise what is meant by the initial state Eq. (\ref{initialstate}). The
indistinguishability of atoms at different sites is also important from a
practical point of view because it allows one to squeeze the 
pseudo-spin for all atoms, not just the atoms at each lattice site.

We expand the atomic field operators in an energy basis $\varphi_n(z)$ of the
single harmonic oscillator representing the lattice,
\begin{equation}
\hat \psi_\sigma(z)=\sum_n\varphi_n(z)\hat c_{\sigma,n},
\end{equation}
with oscillator frequency $ \omega_{\rm osc}=\sqrt{4\hbar
|g_0| |\beta_+|^2 k_L^2/M} $ and oscillator length $a_{\rm
osc}=\sqrt{\hbar/(M \omega_{\rm osc})}$.
We assume that the atoms are deep in the Lamb-Dicke regime which for atoms in
the first few vibrational states means $k_L a_{\rm osc},\; k_c a_{\rm osc}\ll
1.$

\subsection{Mean field equations}

In the rest of this article, we study this system in the mean field
approximation. The mean field equations of motion for the field amplitudes
$\langle \hat d_\pm \rangle$ and the atomic amplitudes $\langle \hat
c_{\sigma,n} \rangle$ are found from the Liouville equation for the density
matrix
\begin{equation}
\frac{d \hat\rho(t)}{dt}=\frac{-i}{\hbar}[\hat H,\, \hat \rho]+\mathcal{L}[\hat\rho].
\end{equation}
These equations close if we factorize the correlations between atoms and light
field according to $\langle \hat d_\pm \hat c_{\sigma,n}
\rangle\rightarrow\langle \hat d_\pm
\rangle\langle \hat c_{\sigma,n} \rangle$, $\langle \hat d_\pm^\dagger
\hat d_\pm\rangle\rightarrow\langle\hat d_\pm^\dagger\rangle\langle
\hat d_\pm\rangle$, etc.
We find
\begin{widetext}
\begin{eqnarray}
\label{eqnen}
i \frac{d \langle \hat c_{e,n}\rangle}{dt}&=&
n\omega_{\rm osc}\langle\hat c_{e,n}\rangle
+\frac{\Omega e^{-i\delta t}}{2}\sum_{n^\prime}\langle n|e^{ik_{c}z}|n^\prime\rangle \langle \hat c_{g,n^\prime}\rangle\\
&&+i\hbar g_0\left((\beta^*_++\langle\hat d_+^\dagger\rangle) \langle \hat d_-\rangle -(\beta_+ +\langle\hat d_+\rangle)\langle \hat d_-^\dagger \rangle\right)\sum_{n^\prime}\langle n|\sin\, 2k_L z|n^\prime\rangle \langle \hat c_{e,n^\prime}\rangle,\nonumber\\
\label{eqngn}
i \frac{d \langle\hat c_{g,n}\rangle}{dt}&=&
n\omega_{\rm osc}\langle\hat c_{g,n}\rangle
+\frac{\Omega e^{i\delta t}}{2}\sum_{n^\prime}\langle n|e^{-ik_{c}z}|n^\prime\rangle \langle \hat c_{e,n^\prime}\rangle\\
&&+i\hbar g_0\left((\beta^*_++\langle\hat d_+^\dagger\rangle) \langle \hat d_-\rangle -(\beta_+ +\langle\hat d_+\rangle)\langle \hat d_-^\dagger \rangle\right)\sum_{n^\prime}\langle n|\sin\, 2k_L z|n^\prime\rangle \langle \hat c_{g,n^\prime}\rangle,\nonumber\\
i\frac{d\langle \hat d_-\rangle}{dt}&=&
-(\Delta-2g_0 \mathcal{S}(t)+i\kappa/2)\langle \hat d_-\rangle
-ig_0(\beta_++\langle\hat d_+\rangle)\mathcal{S}_2(t),\\
i\frac{d\langle \hat d_+\rangle}{dt}&=&
-(\Delta-2g_0 \mathcal{C}(t)+i\kappa/2)\langle \hat d_-\rangle
+ig_0\langle\hat d_-\rangle\mathcal{S}_2(t)
+g_0(\mathcal{C}_2(t)-\mathcal{C}_2(0))\beta_+.
\label{eqnaplus}
\end{eqnarray}
\end{widetext}
We have introduced
\begin{eqnarray}
\mathcal{S}(t)&=&\sum_\sigma \int dz\;\sin^2\,k_L z\langle \hat \psi_\sigma^\dagger(z)\hat\psi_\sigma(z)\rangle ,\\
\mathcal{S}_2(t)&=&\sum_\sigma \int dz\;\sin\,2k_L z\langle \hat \psi_\sigma^\dagger(z)\hat\psi_\sigma(z)\rangle,
\end{eqnarray}
and
\begin{equation}
\mathcal{C}_2(t)=\sum_\sigma \int dz\;\cos\,2k_L z\langle \hat \psi_\sigma^\dagger(z)\hat\psi_\sigma(z)\rangle.
\end{equation}

\subsection{Approximations}

In the Lamb-Dicke and resolved sideband regime the density distribution of the
atoms does not change much during the evolution. In particular we have
\begin{equation}
\mathcal{C}_2(t)-\mathcal{C}_2(0)=\mathcal{O}((k_L a_{\rm osc}\frac{\Omega}{\omega_{\rm osc}})^2).
\end{equation}
Neglecting this small term in Eq. (\ref{eqnaplus}) is an excellent
approximation for the cases we are interested in.

With this approximation it is clear that the modes $\hat d_-$ and $\hat d_+$
are no longer pumped. Light is scattered into these modes exclusively through
interaction with the atoms and we find the scalings
\begin{equation}
\langle \hat d_+\rangle \sim 
\frac{g_0\mathcal{S}_2(t)}{\kappa}\langle \hat d_-\rangle,
    \quad
    \langle \hat d_-\rangle
    \sim
\frac{g_0\mathcal{S}_2(t)}{\kappa}\beta_+.
\end{equation} 
As discussed in the introduction we consider the bad cavity limit. The
previous equation motivates the appropriate condition for the bad cavity limit,
\begin{equation}
\frac{g_0\mathcal{S}_2(t)}{\kappa}\ll 1.
\label{badcavitycondition}
\end{equation}
The resulting hierarchy of the amplitudes $\langle \hat d_+\rangle\ll \langle
\hat d_-\rangle \ll \beta_+$ allows us to make further approximations: We
neglect the symmetric mode altogether, $\langle \hat d_+\rangle\equiv 0$, and we
neglect the back-action of the field on the atoms in Eqs.
(\ref{eqnen},\ref{eqngn}) which is of order $g_0\beta_+ \langle \hat d_-\rangle
$ compared to the lattice potential of order $g_0\beta_+^2$ contained in
$\omega_{\rm osc}$.

We end up with a theory in which the atoms move in the steady state lattice potential and are driven by the clock laser. Through $\mathcal{S}_2(t)$ they are a source for the $\langle \hat d_-\rangle$ field.

\section{\label{IntensityImbalance} Intensity imbalance}

\begin{figure}
\includegraphics[width=8cm]{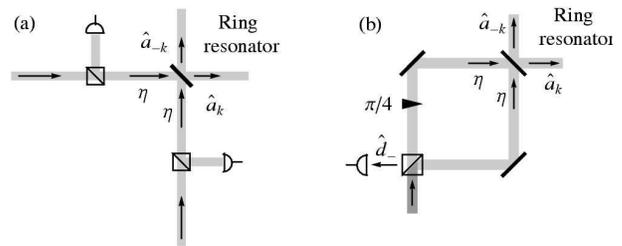}
\caption{In- and out-coupling into the ring cavity as well as detector
    arrangements for the two measurement schemes discussed in the text. In (a)
	two phase-locked lasers pump the two counter propagating modes of the
	ring cavity and the light leaking out of each mode is picked up with
	beam splitters. In (b) an incoming pump beam is sent through a
	Mach-Zehnder interferometer-like arrangement with a $\pi/4$ phase shift
	in one of the arms. This way the symmetric superposition $\hat b_+$
	of the two running wave modes is pumped. The $\hat d_-$ amplitude leaks
	out of the empty port of the Mach-Zehnder interferometer where its
	intensity is measured.}
	\label{detectionschemes}
\end{figure}

As discussed in the introduction our goal is to find the population difference
between electronic excited and ground states after a $\pi/2$-pulse by measuring
the momentum transfer between the $+k_L$ and $-k_L$ modes. A schematic of the detector arrangement for this measurement is shown in Fig. \ref{detectionschemes}(a).

To find the intensity imbalance between the modes we integrate the mean field
equations of motion numerically. From the numerical solutions for the
amplitudes inside the cavity we find the intensity difference at the cavity
output 
\begin{equation}
\delta I=\kappa \langle \hat a_{k_L}^\dagger \hat a_{k_L}-\hat a_{-k_L}^\dagger \hat a_{-k_L}\rangle =2 \kappa {\rm Re}\; \beta_+^*\langle \hat d_-\rangle.
\end{equation}
Of course there will be noise in the measurement of the intensity imbalance. In order to find out whether this QND scheme is suitable for spin squeezing we have to determine whether it is possible to measure the number difference with a precision better than the atomic shot noise despite the noise in the intensity imbalance. 

\begin{figure}
\includegraphics{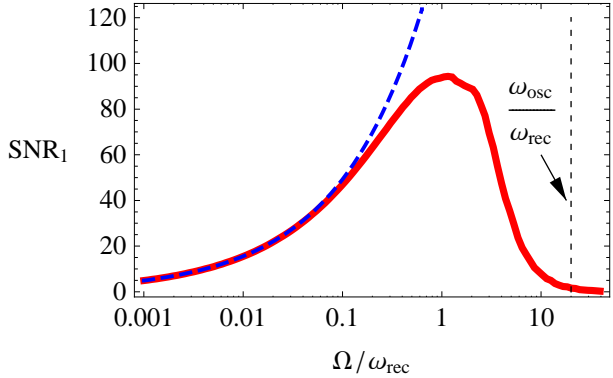}
\caption{(Color online) SNR for detection of the intensity imbalance
during a $\pi/2$-pulse for $10^6$ ${}^{87}$Sr atoms in a ring cavity
of length $L$ = 1 cm, finesse $F=10^6$, and waist $w_0$ = 30 $\mu$m.
$\omega_{\rm osc}=20 \omega_{\rm rec}$ and $\delta=0$ (red solid line).
The lattice laser is tuned to the magic wavelength $\lambda_L=813 nm
$ and the polarizability at that wavelength is $\alpha=-5.37\times
10^{-28} \text{m}^3\epsilon_0$. The blue dashed line indicates the
analytic result (Eq.~(\ref{DCSNRweakdrive})) in the adiabatic
limit.} \label{SNRRingCavity}
\end{figure}

We assume that the detection of the intensities of the two modes is
(photon-) shot noise limited. The photon shot noise 
in each port is $\sqrt{\langle \hat a_{\pm k_L}^\dagger \hat a_{\pm
k_L}\rangle}$
\footnote{We neglect partition noise due to the beam splitters used to pick up
the light leaking out of the two modes in Fig.~\ref{detectionschemes}. This
does not affect our conclusions.}. The resulting SNR can be
calculated as
\begin{equation}
\text{SNR}_1=\frac{\int_0^{\frac{\pi}{2\Omega}}dt \kappa \langle
\hat a_{k_L}^\dagger \hat a_{k_L}-\hat a_{-k_L}^\dagger \hat
a_{-k_L}\rangle}{\sqrt{\int_0^{\frac{\pi}{2\Omega}}dt \kappa \langle
\hat a_{k_L}^\dagger \hat a_{k_L}+\hat a_{-k_L}^\dagger \hat
a_{-k_L}\rangle}}.
\label{SNR1}
\end{equation}
Figure~\ref{SNRRingCavity} shows the result as a function of the drive strength
$\Omega$ of the clock transition for realistic lattice clock parameters. The
clock laser is resonant with the $n=0\rightarrow n=0$ transition, i.e.
$\delta=0$.  The lattice laser amplitude $\eta$ has been adjusted to give
$\omega_{\rm osc}=20 \omega_{\rm rec}$, where $\omega_{\rm rec}=\hbar
\omega_L^2/(2Mc^2)$ is the recoil frequency at the lattice wavelength. We have assumed
that the atoms are driven from ground to excited state with a $\pi/2$-pulse as
they would during the first pulse of a Ramsey sequence.  The detector outputs
are recorded only during the pulse as indicated by formula Eq.
(\ref{SNR1}).

In the resolved side band limit $\Omega \ll \omega_{\rm osc}$ we can
obtain an approximate analytical solution of the mean field equations of motion
by assuming that the coherences between different harmonic oscillator levels
follow the drive adiabatically. From the adiabatic solutions for the atomic
amplitudes we can then in turn find adiabatic solutions for the light field
provided that the strength of the clock laser coupling is well within the
band-width of the cavity, $\Omega\ll \kappa$. Once we have the amplitudes of
the lattice modes we can calculate the signal to noise ratio using Eq.~(\ref{SNR1}). We find
\begin{equation}
{\rm SNR}_1^{\rm ad.}=
\frac{N k_c/k_L}{\sqrt{\frac{\pi  \kappa  \omega_{\rm osc}^2}{16 |g_0| \Omega  \omega_{\rm rec}}}}.
\label{DCSNRweakdrive}
\end{equation}
As can be seen in Fig.~\ref{SNRRingCavity}, the analytical solution
agrees very well with the numerical results in the limit
$\Omega/\omega_{\rm osc}\ll 1$.

The SNR falls off for small $\Omega$ because the signal strength is limited by
a photon redistribution of order $\sim N/2$ while the number of photons that
contribute to the shot noise keeps increasing over longer intervals
$\pi(2\Omega)^{-1}$. At large $\Omega$ the SNR falls off because more and more
momentum is taken up by the atoms as they are driven harder and accordingly less
intensity needs to be redistributed between the lattice modes to ensure
conservation of total momentum. The maximum SNR is obtained near $\Omega\approx
\omega_{\rm osc}$ \footnote{Note that for our parameters the maximum SNR is
obtained closer to $\Omega\approx \omega_{\rm rec}$. For that Rabi frequency
$\sqrt{\omega_{\rm rec}/\omega_{\rm osc}}$ should be replaced by $\omega_{\rm
rec}/\omega_{\rm osc}$, a factor of $1/4$ difference which doesn't affect our
conclusions.} . Assuming that the
expression Eq.~(\ref{DCSNRweakdrive}) still holds, we estimate the maximum SNR
as
\begin{equation}
{\rm SNR}^{\rm max}_{1}=\sqrt{N}\frac{4}{\sqrt{\pi}}\frac{k_c}{k_L}\sqrt{\frac{\omega_{\rm rec}}{\omega_{\rm osc}}}\sqrt{\frac{Ng_0}{\kappa}}.
\label{snrmaxDC}
\end{equation}
This simple scaling law is one of the key results of this paper.
In order for the measurement to lead to spin squeezing, the
$\rm{SNR}$ has to be sufficiently large such that the measurement
uncertainty is smaller than the atomic projection noise, i.e., ${\rm
SNR}>N/\Delta N\sim \sqrt{N}$. In other words the measurement has to be atomic
shot noise limited. Since $4k_c(\sqrt{\pi} k_L)^{-1}(\omega_{\rm rec}/\omega_{\rm
osc})^{1/2}$ is typically of order one and is hard to vary in experiments the
collective
coupling parameter $Ng_0\kappa^{-1}$ is the all important parameter. Note that
we can be in the strong collective coupling regime \cite{Hemmerich} $Ng_0\kappa^{-1}\gg1$
while still in the bad cavity limit in the sense of Eq.
(\ref{badcavitycondition}) because $\mathcal{S}_2(t)\ll N$ in the Lamb-Dicke and
resolved side band regime.

It can be seen from Eq.~(\ref{snrmaxDC}) and Fig.~\ref{SNRRingCavity} that 
a $\rm{SNR}$ greater than $\sqrt{N}$ is hard to achieve with currently
realistic lattice clocks but it might not be completely out of reach in the
future. The collective coupling parameter is small in these systems primarily
because of the requirement of operating the lattice at the magic wavelength.
The magic wavelength is typically very far removed from atomic resonances.
Therefore the atomic polarizability is very small, giving rise to a small single
atom coupling constant $g_0$. The fundamental reason for the rather limited
$\rm{SNR}$ is the intrinsic photon shot noise of the lattice beams, i.e. the
measurement is photon-shot noise limited rather than atom-shot noise limited.
The momentum transfer from the clock laser to the lattice laser that has to be
measured is smaller than the large momentum uncertainties stemming from the
photon shot noise in the lattice beams. In the conclusion section we discuss
methods that may improve the SNR of this scheme to a level where it becomes a
viable means to obtain spin squeezing. However, we will first introduce a much
more promising approach in the next section for measurement-induced spin
squeezing.

\section{\label{SidebandSpectroscopy}Sideband spectroscopy}

In this section we describe a superior detection method that is not
affected by the large photon shot noise in the lattice beams that dominated the
SNR in the detection method discussed in the previous section. The general idea
is to design a measurement in which the signal can be separated from the
lattice beams. The motional sidebands generated by the atoms oscillating in the
lattice with frequencies $\pm\omega_{\rm osc}$ are such a signal. In a
heterodyne detector with bandwidth much smaller than $\omega_{\rm osc}$ the
sidebands can be distinguished from the carrier at $\omega_L$, provided that the
line width of the lattice is smaller than $\omega_{\rm osc}$ which is 
achieved experimentally.

In order for the detection of atomic vibration to constitute a measurement of
the populations of atomic electronic states these two quantities must be
strongly correlated. If for instance only atoms of one electronic state are
oscillating while the other is at rest, detection of the sideband
can measure the number of atoms in the oscillating state. Other possibilities
would be to have atoms of both states oscillate with the same amplitude but
$\pi/2$ or $\pi$ out of phase. If the atoms oscillate $\pi/2$ out of phase the
populations of the two states are proportional to the intensities in the two
quadratures of the sidebands. If they oscillate $\pi$ out of phase, the
intensity of the sidebands is directly proportional to the number difference of
the two states.

Such correlated states of atomic motion and internal level are created rather
naturally in lattice clocks. To see this we consider the atomic dynamics during
a $\pi/2$-pulse starting from initial state Eq. (\ref{initialstate}). The
limit of a strong pulse $\Omega\gg \omega_{\rm osc}$ is easy to understand
intuitively. In this case the component of the atoms that undergoes
a transition to the excited state receives $N_e \hbar k_c$ of recoil momentum 
where $N_e$ is the final population of the excited state, while the atoms that
stay in the ground state remain at rest. After the $\pi/2$-pulse only the
excited state component will oscillate at frequency
$\omega_{\rm osc}$.

\begin{figure}
\includegraphics[width=7cm]{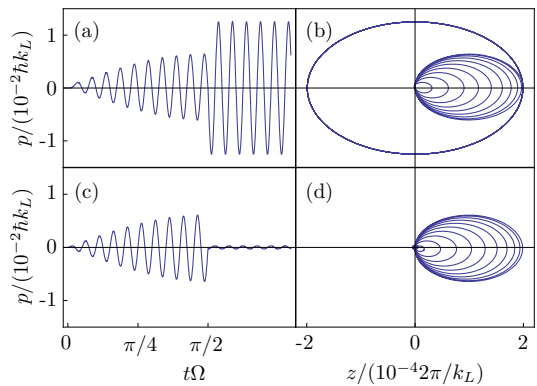}
\caption{(Color online) Momentum per atom in ground (a) and excited state (c)
during a $\pi/2$-pulse with $\Omega=0.5\omega_{\rm rec}$. Figures (b) and (d)
show the corresponding phase space trajectories for atoms in ground and excited
state, respectively. At the beginning of the pulse the atoms are at rest at the
origin, $p=0$ and $z=0$. During the pulse the atoms oscillate on ellipses of
growing diameter. When the pulse terminates at $t=\pi/2/\Omega$ the atoms in
the ground state are at the turning point on the right. Therefore they continue
to oscillate with large amplitude in the harmonic trap after the pulse. The
excited state on the other hand is at the origin at the end of the pulse. It
therefore remains essentially at rest after the pulse.}
\label{phasespaceplot}
\end{figure}

The atomic motion is more complex in the weak drive limit $\Omega\ll
\omega_{\rm osc}$. Because atoms exchange momentum with the lattice lasers
while absorbing and reemitting clock photons it is clear that both states start
oscillating. The atoms' dynamics is illustrated in Fig. \ref{phasespaceplot}.
During the pulse the atoms in the two electronic states oscillate $\pi$ out of
phase with each other but with equal envelope. The amplitude of the
oscillations is suppressed compared to the strong pulse limit by a factor $k_c
a_{\rm osc}\Omega/\omega_{\rm osc}$. If the pulse duration is
$t_{\pi/2}\omega_{\rm osc}=m2\pi$ with $m$ an integer, only the ground state
oscillates after the pulse. For $t_{\pi/2}\omega_{\rm osc}=(m+1/2)2\pi$ only
the excited state oscillates and for $t_{\pi/2}\omega_{\rm osc}=(m\pm 1/4)\pi$
both states oscillate with equal amplitude. The oscillations after the pulse
are undamped to a good approximation for $\kappa\gg \omega_{\rm osc}$ and
$\Delta=2g_0N$ regardless of the strength of $\Omega$ \cite{Domokos:Mechanical_effects_light,Horak:Dissipative_BEC_cavity}.

For the numerical examples in this section we consider a cavity with reduced
finesse $F=10^4$ (corresponding to $\kappa\approx 2\pi\; (3\; {\rm MHz})$ for $
L=1$ cm) compared to the example in the previous section to ensure that
$\kappa\gg \omega_{\rm osc}$. All other parameters are as before. 

We now turn to a quantitative discussion of the measurement of the atomic
vibration using the lattice fields. As before we assume that we are in the bad
cavity limit Eq.  (\ref{badcavitycondition}) and we retain $\langle \hat
d_-\rangle$
and neglect $\langle \hat d_+\rangle$.  We consider the situation outlined
above where $t_{\pi/2}\omega_{\rm osc}=m2\pi$ such that the excited state is at
rest after the $\pi/2$ pulse. The amplitude $\langle \hat d_-\rangle$ is then
fed by a source
\begin{equation}
g_0\sum_{n,n^\prime,\sigma}\langle n|\sin\, 2k_Lz|n^\prime\rangle \langle \hat c_{n,\sigma}^\dagger\rangle\langle\hat c_{n^\prime,\sigma}\rangle\approx g_0 N_g\bar{z}_g\sin(\omega_{\rm osc}t-\phi_g),
\end{equation}
where $N_g$ is the population of the ground state and $\bar{z}_g$ and $\phi_g$ are the amplitude and phase of the oscillations of the ground state atoms. 

\begin{figure}
\includegraphics[width=7cm]{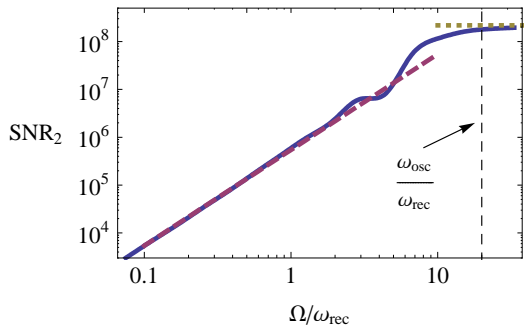}
\caption{(Color online) SNR for 1 s of detection at the vibrational
sideband for ${}^{87}$Sr. The asymptotic
solutions for weak and strong drives, Eqs.~(\ref{SNR2weak}) and
(\ref{SNR2max}), are shown as the dashed and dotted lines,
respectively. Parameters are as in Fig. \ref{SNRRingCavity} except that
    the cavity finesse is $F=10^4$.
}\label{SNRRingCavityAC}
\end{figure}

In order to evaluate the suitability of this scheme for spin squeezing we need
to calculate the SNR for measuring the intensity of the sideband $\langle \hat
d_-\rangle$ \footnote{Note that the signal $\langle \hat d_-\rangle$ is
    separated from the carrier $\beta_+$ not only in frequency but also
	interferometrically, see Fig.~\ref{detectionschemes}; $\mathcal{S}_2(t)=0$ if the atoms are at rest and
	no light leaks out of the $\hat d_-$ mode.}. For heterodyne detection 
with a strong local oscillator the SNR is
\begin{equation}
{\rm SNR_2}=\int_{t_{\pi/2}}^{t_{\pi/2}+T}dt \kappa |\langle \hat d_-\rangle|^2,
\end{equation}
where $T$ is the integration time and we start the measurement immediately
after the $\pi/2$-pulse. With a strong local oscillator every $\hat d_-$ photon
is detected and the SNR is equal to the total number of photons scattered from
the symmetric mode into the $\hat d_-$ mode. The SNR can be readily evaluated
using the numerical solutions for the field amplitude. The result is shown in
Fig. \ref{SNRRingCavityAC} for an integration time of 1 s as a function of
$\Omega$. As in the previous section the weak drive limit can be studied
analytically by assuming that the coherences between the different atomic
vibrational levels follow the drive adiabatically. The adiabatic result,
\begin{equation}
{\rm SNR}_2^{\rm ad.}=\frac{(k_c/k_L)^2 N^2 \Omega^2 \kappa
T}{32(\Delta^2+(\kappa/2)^2)|\beta_+|^2},
    \label{SNR2weak}
\end{equation}
is also shown in Fig.~\ref{SNRRingCavityAC} and agrees well with the numerical
result. The amplitude of the atoms' oscillations decreases as the strength of
the drive is reduced according to the aforementioned suppression by
$\Omega/\omega_{\rm osc}$, resulting in a weaker generated signal and accordingly a smaller SNR.

In the strong drive limit the atoms eventually take up all the recoil momentum
$(N/2)\hbar k_c$ and the SNR saturates at
\begin{equation}
{\rm SNR}_2^{\rm max}=\frac{(k_c k_L a_{\rm osc}^2)^2 N^2 |\beta_+|^2 g_0^2 \kappa T}{2(\Delta^2+(\kappa/2)^2)}.
\label{SNR2max}
\end{equation}

Figure~\ref{SNRRingCavityAC} shows that for $\Omega$ $\sim$
$\omega_{\rm rec}$, a SNR $>$$10^5$ can be achieved. For $10^6$
atoms, this corresponds to a measurement uncertainty a hundredfold
smaller than the projection noise, indicating that spin squeezing
becomes possible with this measurement scheme. Note that even at
$\Omega\sim\omega_{\rm rec}$, the population of the first excited
vibrational state is suppressed by a factor of $(k_c a_{\rm osc}
(\Omega/\omega_{\rm osc}))^2\approx 10^{-4}$ relative to the
vibrational ground state, thus maintaining the validity of the Lamb-Dicke
regime.

\section{\label{Conclusion}Conclusion}

We have studied the feasibility of measuring the populations of clock states
using the lattice field of an optical lattice clock, with the goal of squeezing
the atomic pseudo-spin. We have demonstrated that a measurement of population
transfer based on detection of momentum redistribution between the lattice
beams is possible but, with current experimental technology, not sensitive
enough to enable spin squeezing.

The reason for this failure is that the collective atom lattice coupling
$Ng_0/\kappa$ is too small for a lattice operating at the magic wavelength. A
possible solution might be to operate the lattice at other magic frequencies
where $g_0$ is larger. Engineering the magic wavelength with external electric,
magnetic or laser fields such that it is closer to an atomic resonance might be
another option. Stronger coupling can also be achieved with a smaller volume
cavity.

A second measurement scheme based on detecting the motional sidebands of atoms
after a clock pulse is promising. The SNR that we calculated for realistic
parameters appropriate for a ${}^{87}{\rm Sr}$ lattice clock suggests that spin
squeezing could be feasible with this method.

Several questions will have to be addressed in the future. First it will be
important to go beyond the harmonic approximation in which the optical lattice
is treated as a single harmonic trap. Among other things this will allow us to
elucidate the role of tunneling and the band structure during and after the
clock pulses. Second the system needs to be studied in the good cavity limit.
The good cavity limit is particularly interesting since our results indicate
that the two detection schemes presented here work better in cavities with
higher $Q$-factor.

Already in the bad cavity, but especially in the good cavity case, it will be
necessary to go beyond the mean field approximation. That would allow one to
study in detail how the noise in one of the pseudo spin components is reduced
at the expense of the other components. Such a beyond meanfield treatment is
indispensable for finding the ultimate resolution limits of the spin
measurements presented here. Possible routes are cumulant expansion of the
atomic and lattice field operators, Langevin equations and Monte-Carlo
wavefunction methods.

A question of fundamental interest arising in the context of the second
measurement scheme is that in that case we are trying to measure states that
are neither eigenstates of the Hamiltonian nor of the operator $\sin 2k_L z$
with which the light field couples to the atoms. Instead we are trying to
differentiate between states that differ in their dynamics, i.e. one of them is
oscillating while the other is at rest. A measurement can therefore not be done
instantaneously since this would project the system on eigenstates of $\sin
2k_L z$. Rather one has to carefully erase all information about the time at
which a photon has been scattered from the symmetric mode into the
antisymmetric mode by having the photons circulate in the cavity for a
sufficiently long time before they leak out.

We also plan to investigate the prospects of quantum feedback control
\cite{JMGeremia04092004,Thomsen} that should allow one to not only prepare the
many-particle state probabilistically in squeezed states with a certain $J_z$
projection, but also to deterministically drive the system to a target $J_z$
projection.

We acknowledge fruitful discussions with H. J. Kimble, P. Jessen, M.
M. Boyd, A. Ludlow, and H. Ritsch. This work has been supported by DOE, NIST,
and NSF. D. M. gratefully acknowledges support from DAAD. His email address is
dmeiser@jila.colorado.edu.

\bibliography{mybibliography}

\end{document}